\documentclass[11pt,twoside]{article}

\usepackage{asp2014}

\aspSuppressVolSlug
\resetcounters

\begin{document}

\title{(the struggle) Towards an open source policy}

\author{Y.~G.Grange$^1$, T. J\"urges$^1$, T~.J. Dijkema$^1$, R. Halfwerk$^{1,2}$, and G.~W. Schoonderbeek$^1$}
\affil{$^1$ASTRON, the Netherlands Institute for Radio Astronomy, Oude Hoogeveensedijk 4, 7991 PD, Dwingeloo, The Netherlands; \email{opensource@astron.nl}}
\affil{$^2$AstroTec Holding, Oude Hoogeveensedijk 4, 7991 PD, Dwingeloo, The Netherlands}

\paperauthor{Y.~G. Grange}{grange@astron.nl}{0000-0001-5125-9539}{ASTRON}{Astronomy Group}{Dwingeloo}{}{7991PD}{The Netherlands}
\paperauthor{T. J\"urges}{jurges@astron.nl}{0000-0002-3993-7576}{ASTRON}{Radio Observatory}{Dwingeloo}{}{7991PD}{The Netherlands}
\paperauthor{T.~J. Dijkema}{dijkema@astron.nl}{}{ASTRON}{Research \& Development}{Dwingeloo}{0000-0001-7551-4493}{7991PD}{The Netherlands}
\paperauthor{R. Halfwerk}{halfwerk@astron.nl}{}{AstroTec Holding}{}{Dwingeloo}{}{7991PD}{The Netherlands}
\paperauthor{G.~W. Schoonderbeek}{schoonderbeek@astron.nl}{0000-0001-9482-1253}{ASTRON}{Research \& Development}{Dwingeloo}{}{7991PD}{The Netherlands}


\begin{abstract}
Public availability and tracability of results from publically-funded work is a topic that gets more and more attention from funding agencies and scientific policy makers. However, most policies focus on data as the output of research. In this contribution, we focus on research software and we introduce the ASTRON Open Source Policy. Apart from the license used (Apache 2.0), the policy is written as a manual that explains how to license software, when to assign a Digital Object Identifier (DOI), and defines that all code should be put in an ASTRON managed repository. The policy has been made publically available, a DOI has been assigned to it and it has been put in a repository to stimulate the ADASS community to start a conversation on how to make our code publically accessible and citable. 
\end{abstract}


\section{Introduction}
In recent years, the idea that any result that has been created based on research funded by public money, should be publically available has gained momentum (e.g. \citealp{2016_Wilkinson_FAIR}). Most policies that have been created in this context are however focused on the act of making data public.

Source code, that for instance reduces or creates data, on the other hand differs from data very much because the goal of the code is neither to be read nor to be analysed but to be executed and hence per definition remain invisible.

	In theory this does not pose a big problem. If all source code is licensed under a well-known open source license that adhere to the definition of open source that was defined by the Open Source Initiative (OSI) \footnote{https://opensource.org/docs/osd}, then one should think that all is good and done.  But in reality licensing of source code in scientific or research projects that are publicly funded faces several issues:
\begin{enumerate}

\item \label{it:quote} Accountability for the results by visibility of the implementation of algorithms and other source code are still not a common theme in the scientific community. 
\item \label{it:open} Publicly funded projects that produce source code have to release the source code as open source.
\item \label{it:recog} There are no uniform policies for public recognition of the source code development work, like there are for the work of performing scientific research.  
\item \label{it:ideas} \label{it:last} A societal call for profit-building use of research appears to work against the \textit{ideas} of open source code.
\end{enumerate}

\section{Open Source Code and Reality:  Opponents or a Chance?}
    This leads to the question if it is possible to solve the above mentioned issues: accountability (\#\ref{it:quote} above), availability (\#\ref{it:open}), recognition (\#\ref{it:recog}) and yielding a profit(\#\ref{it:ideas}) by simply picking one of the widely-available open source licenses.
    Some open source licenses contain clauses on patents (e.g. BSD+Patent \footnote{https://opensource.org/licenses/BSDplusPatent}), some emphasise on openness and availability of source code (e.g. GPL\footnote{https://www.gnu.org/licenses/}) but it appears that no commonly known open source license exists that makes the origins and ideas behind derived code explicitly visible.
    
    Creating a new, OSI compliant, license that fits exactly the needs of our field would need significant legal input from our organisation to make sure the license is legally valid. However, this would also require a legal check by the legal department of any collaborating entity or organisation wishing to extend the code, produced by us, to assess if our license fits their corporate policies. In practise this will result in outright rejection of the license, and therefore severely limit the code utilisation compared to code using a well-known established open source license.

Hence it is obvious that choosing the right license for source code that is developed for a project is not simple. Not only is the vast amount of already existing licenses making the right choice complicated, but also the encouraging of every software developer in the project to apply it can be even more of a challenge.  This is the point where a step-by-step instruction manual that explains how to deal with the aspects of open sourcing project source code would come handy. In management speak one could also refer to this manual as a \textit{policy}.

\section{ASTRON's Open Source Committee}
The Netherland Institute for Radio Astronomy (ASTRON) recognised the need for an Open Source Policy.  It installed a committee that was tasked to evaluate commonly known and widely used open source licenses, to choose one, and finally to write a policy on this matter.
	The committee consists of members from each department within ASTRON that develops software. The composition of the committee at the time of writing coincides with the author list of this paper. The first action of the committee was to set up its terms of reference \citep{2017_ASTRONOSC_TOR}. 

\section{ASTRON's choice of License}
The Open Source Committee chose the Apache~2.0\footnote{https://apache.org/licenses/LICENSE-2.0} license as the base for ASTRON's future Open Source Policy. The choice was influenced by the request of the Dutch organisation for scientific research (NWO) to make all results of research accessible without any legal restrictions. Applying a license automatically imposes legal restrictions (paradoxically not applying a license to code makes it protected by copyright which in itself restricts usage as well), the committee chose one of the licenses that impose as little constraints as possible on the re-use of source code. 

\section{ASTRON's Open Source Policy}
ASTRON's Open Source Committee has tried to write the Open Source Policy in such a way that it is independent of the exact choice of license. Also, the goal of the written document is not only to be a policy document, but also, as mentioned earlier, act as a manual for project leads and developers on the application of the license for their work. 

In several cases, like projects that are part of a collaboration, or projects that are funded by grants which have specific requirements on licensing, it may be impossible to apply the policy. The Open Source Committee has been mandated by the management to judge such cases. The main philosophy of the committee is to apply as many components of the policy as possible in those cases.
The main ingredients of the policy are:

\begin{itemize}
    \item The default license for source code that ASTRON holds a copyright for is Apache 2.0
    \item For each release of the software, a Digital Object Identifier (DOI; \citealp{ISOSTANDARD_DOI}) should be released to support reproducibility of results and citation of the code. We advise to use \citet{ZENODO_DOI}. On top of that, registering code with the Astronomy Source Code Library \citep[ASCL;][]{2015Allen} provides more visibility in our field and is therefore also strongly advised.
    \item The DOI should be put in a file called NOTES which also contains a request for citation. The NOTES file is protected by the Apache 2.0 license, which means that any derived code that is distributed shall contain this file, or an extended version of it.
    \item All software shall be put in a repository that is owned, hosted and managed by ASTRON, making it possible to have a clear overview of all code produced by ASTRON employees.
    \item Any deviations from the policy shall be discussed with the Open Source Committee. The committee will advise ASTRON's management on how to proceed. 
\end{itemize}

\section{Open sourcing the Open Source Policy}
The main idea behind open source software of making source code publicly available has many advantages. First, it offers external parties the opportunity to improve 3rd party software. Further it widens the potential impact of software because the source code is much easier to find and it is free to use for everybody. Lastly, opening the source code offers other developers the opportunity to take (parts of) existing code and incorporate it in their work, increasing the impact of the re-used source code. 

Since all those arguments are equally valid for a policy document, and since the whole point of such a document is to make results of publicly funded work available to the public, we decided to releaae ASTRON's open source policy to the public \citep{2018_ASTRONOSC_OSP}. It has been released under the Creative Commons license (CC BY 4.0), was put on Zenodo and had there a DOI assigned to it. We also have put the policy text in the ASTRON git repository\footnote{https://git.astron.nl/open-source-committee/open-source-policy} (again, as required by the license for source code). We invite all interested parties to use the issue tracker of our repository for feedback and questions\footnote{The ASTRON git server allows for users to log in with their gitlab account.}. 

\section{Towards a repository of policies}
The ASTRON Open Source Committee asks all parties that use ASTRON's policy as input for their own policy to cite it by its DOI (and/or cite this paper). Also, we strongly suggest that any open source policy created by other parties shall be put on Zenodo and be added to the ``Open Source Policy'' community so that this initial idea will grow into a repository that can be used as inspiration for newcomers.

\bibliography{O3-6}  

\acknowledgements The ASTRON Open Source Committee wishes the thank its previous members: Ruud Beukema, Ronald Nijboer, Andr\'{e} Gunst, and Michael Wise for their role in setting up the committee and writing the corresponding policies. We want to thank Lourens Veen from the Netherlands eScience Centre for his feedback and discussions during the early stages of the development of our policy. Also, we are grateful to the management team of ASTRON for their support, and endorsement of the ASTRON Open Source Policy. 

\end{document}